\documentstyle[12pt]{article}
\advance \topmargin by -\headheight
\advance \topmargin by -\headsep
\evensidemargin \oddsidemargin
\title{Gravitational Theory without the 
Cosmological Constant Problem, Symmetries of Space-Filling Branes and 
Higher Dimensions} 
\author{E.I.Guendelman\thanks{GUENDEL@BGUmail.BGU.AC.IL} and
        A.B.Kaganovich \thanks{ALEXK@BGUmail.BGU.AC.IL}}
\date {Physics Department, Ben Gurion University of the Negev, 
   Beer  Sheva 84105, Israel}

\begin{document}
\maketitle
\begin{abstract}
We showed that the principle of nongravitating vacuum energy, when 
formulated in the first order  formalism, solves the cosmological 
constant problem. The most appealing formulation of the theory displays a 
local symmetry associated with the arbitrariness of the measure of 
integration. This can be motivated by thinking of this theory as a direct 
coupling of physical degrees of freedom with a "space - filling brane" 
and in this case such 
local symmetry is related to space-filling brane gauge invariance. The 
model is formulated in the first order formalism using the metric
$G_{AB}$ 
and the connection $\Gamma^{A}_{BC}$ as independent dynamical 
variables. An 
additional symmetry (Einstein - Kaufman symmetry) allows to eliminate the 
torsion which appears due to the 
introduction of the new measure of integration. The most successful  
model that implements these ideas is realized in a six or higher 
dimensional space-time. The compactification of extra dimensions into a 
sphere gives the possibility of generating scalar masses and potentials, 
gauge fields and fermionic masses. It turns out that remaining four 
dimensional space-time must have effective zero cosmological constant.

\end{abstract}

\pagebreak

\section{Introduction}

\bigskip

We have developed a theory \cite{GK1},\cite{GK2} where the measure of 
integration in the action principle is not necessarily $\sqrt{-G}$ ($G=
Det(G_{AB})$) 
but it 
is determined dynamically through additional degrees of freedom. This 
theory is based on the demand that such measure respect  the principle of 
non gravitating vacuum energy (NGVE principle) which states that the 
Lagrangian density $L$ can be changed to $L+constant$ without affecting 
the dynamics. This requirement is imposed in order to offer a new 
approach for the solution of the cosmological constant problem \cite{NWN}.

Clearly the invariance $L\longrightarrow L+constant$ for the action is 
achieved if the measure of integration in the action is a total 
derivative, so that 
to an infinitesimal hypercube in D-dimensional space-time $x_{0}^{A}\leq 
x^{A}\leq x_{0}^{A}+dx^{A}$, $A=0,1,\ldots,D-1$ we associate a volume 
element $dV$ which is: (i) a total derivative, (ii) it is proportional to 
$d^{D}x$ and (iii) $dV$ is a general coordinate invariant. The usual choice, 
$\sqrt{-G} d^{D}x$ does not satisfy condition (i).

All of the conditions (i)-(iii) are satisfied \cite{GK1},\cite{GK2} if 
the measure appropriate to the integration in the space of $D$ 
scalar fields  $\varphi_{a}, (a=1,2,...D)$, that is 
\begin{equation}
dV = 
d\varphi_{1}\wedge 
d\varphi_{2}\wedge\ldots\wedge 
d\varphi_{D}\equiv\frac{\Phi}{D!}d^{D}x
\label{dV}
\end{equation}
where
\begin{equation}
\Phi \equiv \varepsilon_{a_{1}a_{2}\ldots a_{D}}
\varepsilon^{A_{1}A_{2}\ldots A_{D}}
(\partial_{A_{1}}\varphi_{a_{1}})
(\partial_{A_{2}}\varphi_{a_{2}}) \ldots
(\partial_{A_{D}}\varphi_{a_{D}}).
\label{Fi}
\end{equation}

Notice that this is a particular realization of the coupling of $p$-brane 
(with $p+1=D$) with the $(p+1)$-form potential
\begin{equation}
A_{A_{1}A_{2}\ldots A_{D}}=\partial_{[A_{1}}A_{A_{2}\ldots A_{D}]}
\label{ap}
\end{equation}
and a further coupling with the Lagrangian density (usually not 
considered). In fact, if $A_{A_{2}\ldots A_{D}}$ equals to 
$A^{(\varphi)}_{A_{2}\ldots A_{D}}\equiv\frac{1}{D!}\varphi_{a_{1}}
(\partial_{A_{2}}\varphi_{a_{2}}) \ldots
(\partial_{A_{D}}\varphi_{a_{D}})\varepsilon_{a_{1}a_{2}\ldots
a_{D}}$, then $\partial_{[A_{1}} A_{A_{2}\ldots
A_{D}]}dx^{A_{1}}\wedge dx^{A_{2}}\wedge\ldots\wedge dx^{A_{D}}$ 
coincides with (\ref{dV}),(\ref{Fi}).

Following the Ref.\cite{Pol} we will call the brane a "space-filling 
brane" if the rank of the form $A_{A_{1}A_{2}\ldots 
A_{D}}$ that couples to the p-brane equals to the space-time 
dimensionality (that is $p+1=D$). In the normal formulation of $p$-branes 
one requires invariance under gauge transformations of the form
\begin{equation}
A_{A_{1}A_{2}\ldots A_{D}}\longrightarrow A_{A_{1}A_{2}\ldots A_{D}}+
\partial_{[A_{1}}\Lambda_{A_{2}\ldots A_{D}]}
        \label{ALamb}
\end{equation}
and simply write the coupling $g\int A_{A_{1}A_{2}\ldots 
A_{D}}dx^{A_{1}}\wedge dx^{A_{2}}\wedge\ldots\wedge dx^{A_{D}}$ which is 
invariant under 
(\ref{ALamb}) provided $\Lambda_{[A_{2}\ldots A_{D}]}\rightarrow 0$ as 
$x^{A}
\rightarrow\infty$ and one doesn't allow coupling to a third entity (like 
the Lagrangian density $L$).

The problem is\cite{Pol} that  in the case of 
a "space-filling brane" the equation of motion obtained from 
varying $A_{A_{1}A_{2}\ldots
A_{D}}$ is simply $g=0$, that is  there is no action principle to talk 
about.

In the alternative we propose, we don't have to necessarily insist on the 
particular realization (\ref{dV}),(\ref{Fi}), although it has the most 
attractive geometrical interpretation\cite{GK2}. 

We will choose to write $A_{A_{1}A_{2}\ldots A_{D}}$ as a total derivative
\begin{equation}
A_{A_{1}A_{2}\ldots A_{D}}=
\partial_{[A_{1}}A_{A_{2}\ldots A_{D}]}
        \label{PG}
\end{equation}
and then we use $A_{A_{2}\ldots A_{D}}$ as the independent dynamical 
variables in our action principle. Furthermore we can implement the 
NGVE-principle if we write the following action
 \begin{equation}
S=\int L\partial_{[A_{1}}
A_{A_{2}\ldots A_{D}]}dx^{A_{1}}\wedge dx^{A_{2}}\wedge\ldots\wedge 
dx^{A_{D}}
    \label{Act}
\end{equation}
which describes the coupling of the brane to gravity and matter which 
appear through the Lagrangian density $L=L_{g}+L_{m}$. The Lagrangian 
structure has to be defined by the demand that the action (\ref{Act}) be 
invariant under the following gauge transformation 
\begin{equation}
A_{A_{2}\ldots A_{D}}\longrightarrow A_{A_{2}\ldots A_{D}}+
\Lambda_{A_{2}\ldots A_{D}}
    \label{BS}
\end{equation}
for any $\Lambda_{A_{2}\ldots A_{D}}$ (without a condition for 
$\Lambda_{A_{2}\ldots A_{D}}$ as $x^{A}\rightarrow\infty$) which will be 
refered afterwards as a "space-filling brane gauge transformation".  In this 
case $L$ has to transform correspondingly in order to compensate the 
transformation of the measure.  How this is realized , will be 
explained after we
understand the basic structure of the theory in the first order 
formalism
(see also Ref.\cite{GK2}).  In the case that we use the representation
(\ref{dV}),(\ref{Fi}), an arbitrary change of the measure corresponds to an 
arbitrary diffeomorphism in the internal space of the scalar fields 
$\varphi_{a}$.

There are two well known variational principles: the
first and the second order formalisms, 
which are equivalent in the case of the general theory of relativity. 
However, 
as we will see, they are inequivalent in our case. In the first order 
formalism , in the action $G_{AB}$ and $\Gamma^{A}_{BC}$ appear, while no 
explicit derivatives of $G_{AB}$ are introduced in the Lagrangian density.
The action principle allows then to solve $\Gamma^{A}_{BC}$ as a function 
of $G_{AB}$ and its first derivatives. The resulting equations are the 
usual Einstein equations which are also obtained from the second order 
formalism which does not involve $\Gamma^{A}_{BC}$ as a dynamical 
variable but rather involves only $G_{AB}$ and their first and second 
derivatives.

In the case at hand, that is in the context of the NGVE theories, the 
first and second order formalisms are not equivalent. The model that 
results from studying the theory in the second order formalism\cite{GK1} 
gives rise to empty space solutions with arbitrary constant curvature.
 In this case the cosmological constant 
problem is not solved (although arguments based on maximal symmetry can 
be made in favor of the zero curvature choice for vacuum). In contrast, 
 the first order formalism leads to the solution of the cosmological 
constant problem in a straightforward way.

In the first order formalism, the theory has been studied \cite{GK2} 
using  the vielbein $e^{A}_{i}$ and the spin-connection 
$\omega^{A}_{ik}$  ($i,k$ denote Lorentz indexes in D dimensions),  
instead of utilizing $\Gamma^{A}_{BC}$ which will be 
the case here. Furthermore, the use of $\Gamma^{A}_{BC}$ as  dynamical 
variables instead of $\omega^{A}_{ik}$ makes manifest a new symmetry of 
the theory, which was discovered as a symmetry of the curvature tensor in 
the affine connection space by Einstein and Kaufman long time ago\cite{E} 
and given by them the name of "$\lambda -transformation$". Although the 
$\lambda$-symmetry was discussed in Ref.\cite{E} in the context of a  very 
specific unified model, it turns out that the range of applicability of 
this symmetry is much wider. This question will be discussed in Sec.2.

The importance of Einstein-Kaufman $\lambda$-symmetry in our model is 
that it allows for the elimination of the torsion in the absence of 
fermions, as opposed with the first order formalism employing 
$\omega^{A}_{ik}$ where it is hard to avoid explicitly the appearence of 
the torsion even in the absence of fermions\cite{GK2}.

In our previous paper \cite{GK2} it was shown that in the first order 
formalism, the theory based on the NGVE-principle possesses an additional 
local symmetry for the vacuum and for some special models. When realizing 
the NGVE-principle with the measure of the form of 
eqs.(\ref{dV}),(\ref{Fi}), we have seen\cite{GK2} that this local symmetry 
incorporates the group of diffeomorphism transformations of the internal 
space of scalar fields $\varphi_{a}$. Here  we will  see that this local 
symmetry can be formulated in a way where it incorporates 
space-filling brane gauge invariance (\ref{BS})(see  sections 3-5 of 
this paper). The importance of this symmetry, apart from its obvious 
geometrical meaning, consists of the fact that for models where it 
holds it is possible to choose the gauge where the measure $\Phi$    
coincides with the measure of general relativity $\sqrt{-G}$. This is 
why we call this symmetry {\em "local Einstein symmetry"}. In Sec.6 
we construct realistic models (without loosing the solution of the 
cosmological constant problem in four dimensions) where the local Einstein 
symmetry holds as an exact symmetry.

\bigskip

\section{Action and Einstein-Kaufman $\lambda$-symmetry}

\bigskip

According to the NGVE-principle, the total action in the D-dimensional 
space-time should be written in the form
\begin{equation}
S=\int\Phi Ld^{D}x
    \label{Act1}
\end{equation}
where 
$\Phi d^{D}x$ may be given either by 
\begin{equation}
\Phi d^{D}x=\partial_{[A_{1}}
A_{A_{2}\ldots A_{D}]}dx^{A_{1}}\wedge dx^{A_{2}}\wedge\ldots\wedge 
dx^{A_{D}}
    \label{MesA}
\end{equation}
(as in eq.(\ref{Act})), or by using $\Phi$ as in  eqs.(\ref{dV}),(\ref{Fi}).
 
We assume that $L$ does not contain the measure fields , that is the 
fields by means of which $\Phi$ is defined. If this condition is 
satisfied then the theory has an additional symmetry. In fact, for 
example for the case of the action with $\Phi$ given by  eq.(\ref{Fi}), the 
action (\ref{Act1}) is invariant under the infinitesimal shift of the 
fields $\varphi_{a}$ by an arbitrary infinitesimal function of the total 
Lagrangian density $L$, that is\cite{GK1},\cite{GK2}
\begin{equation}
\varphi^{\prime}_{a}=\varphi_{a}+\epsilon g_{a}(L),\  \epsilon\ll 1
        \label{LP}
\end{equation} 

Our choice for the total Lagrangian 
density is
\begin{equation}
L=-\frac{1}{\kappa}R(\Gamma,G)+L_{m}
\label{L1}
\end{equation}
where $L_{m}$ is the matter Lagrangian density and $R(\Gamma,G)$ is the 
scalar  curvature 
\begin{equation}
R(\Gamma,G)=G^{AB}R_{AB}(\Gamma)
\label{R}
\end{equation}

\begin{equation}
R_{AB}(\Gamma)=R^{C}_{ABC}(\Gamma)
\label{RAB}
\end{equation}

\begin{equation}
R^{A}_{BCD}(\Gamma)\equiv \Gamma^{A}_{BC,D}-\Gamma^{A}_{BD,C}+
\Gamma^{A}_{ED}\Gamma^{E}_{BC}-\Gamma^{A}_{EC}\Gamma^{E}_{BD}
\label{RABCD}
\end{equation}

 The curvature tensor is invariant under the  
$\lambda$- transformation 

\begin{equation}
\Gamma^{\prime A}_{BC}=\Gamma^{A}_{BC}+\delta^{A}_{B}\lambda,_{C}
\label{Gamal}
\end{equation}
which was discovered by Einstein and Kaufman\cite{E}. Although this 
symmetry was discussed in Ref.\cite{E} in the very specific unified 
theory, it turns out that $\lambda$-symmetry has a wider range of 
validity and in particular it is useful in our case. 

In fact, for a wide class of matter models, the matter Lagrangian density 
$L_{m}$ is invariant under the $\lambda$ transformation too. This is 
obvious if $L_{m}$ does not include the connection $\Gamma^{A}_{BC}$ at 
all (like, for example, for scalar fields, for a point particle and other 
cases that we will discuss in this paper). As an example of particular 
importance we consider  here the case of Dirac fermions in 4-dimensional 
space-time with the hermitian Lagrangian density   
\begin{equation}
L_{f}=-\frac{i}{2}[(\nabla_{\mu}\overline{\psi})\gamma^{\mu}\psi-
\overline{\psi}\gamma^{\mu}\nabla_{\mu}\psi +2iV(\overline{\psi}\psi)]
\label{Lf}
\end{equation}
which is also invariant under $\lambda$-transformation. Here matrices 
$\gamma^{\mu}$ ($\mu=0,1,2,3$) are defined according to 
$\gamma^{\mu}=e^{\mu}_{n}\gamma^{n}$, 
where $\gamma^{n}$ are the Dirac matrices and $e^{\mu}_{n}$ are vielbeins: 
$G^{\mu\nu}=e^{\mu}_{n}e^{n\nu}$. The covariant derivatives in (\ref{Lf}) 
are given by 
$\nabla_{\mu}\psi=\partial_{\mu}\psi-
\frac{1}{4}\Gamma_{\mu\nu\lambda}\gamma^{\nu}\gamma^{\lambda}\psi$,
  $\nabla_{\mu}\overline{\psi}=\partial_{\mu}\overline{\psi}-
\frac{1}{4}\Gamma_{\mu\nu\lambda}\overline{\psi}\gamma^{\nu}
\gamma^{\lambda}$  
and $\Gamma_{\mu\nu\lambda}=G_{\lambda\sigma}\Gamma^{\sigma}_{\mu\nu}$.

What concerns with vector bosons, we note that the demand of gauge 
invariance leads to a generally coordinate invariant gauge boson Lagrangian 
which does not include the connection\cite{HR}.

\bigskip

\section{Connection and local symmetries}

\bigskip

First consider here the case where $L_{m}$ does not depend on 
$\Gamma^{A}_{BC}$, that is fermions and curvature are not present in 
$L_{m}$. 
Varying the action (\ref{Act1}),(\ref{L1}) with respect to 
$\Gamma^{A}_{BC}$,  we get 
\begin{eqnarray}
-\Gamma^{A}_{BC}-\Gamma^{D}_{EB}G^{EA}G_{DC}+\delta^{A}_{C}\Gamma^{D}_{BD}
+\delta^{A}_{B}G^{DE}\Gamma^{F}_{DE}G_{FC}-
\nonumber\\
G_{DC}\partial_{B}G^{DA}+\delta^{A}_{B}G_{DC}\partial_{E}G^{DE}-
\delta^{A}_{C}\frac{\Phi,_{B}}{\Phi}+\delta^{A}_{B}\frac{\Phi,_{C}}{\Phi}
=0
\label{GAM1}
\end{eqnarray}

We will look for the solution (up to a $\lambda$-symmetry transformation) 
of the form
\begin{equation}
\Gamma^{A}_{BC}=\{ ^{A}_{BC}\}+\Sigma^{A}_{BC}
\label{GAM2}
\end{equation}
where $\{ ^{A}_{BC}\}$  are the Christoffel's connection coefficients. Then 
$\Sigma^{A}_{BC}$ satisfies equation
\begin{equation}
-\sigma,_{C}G_{AB}+\sigma,_{A}G_{BC}-G_{BD}\Sigma^{D}_{CA}-
G_{AD}\Sigma^{D}_{BC}+G_{AB}\Sigma^{D}_{CD}+
G_{BC}G_{DA}G^{EF}\Sigma^{D}_{EF}=0
\label{S1}
\end{equation}
where 
\begin{equation}
\sigma\equiv\ln\chi, \hspace{1.5cm} \chi\equiv\frac{\Phi}{\sqrt{-g}}
\label{ski}
\end{equation}

The general solution of eq.
(\ref{S1}) is
\begin{equation}
\Sigma^{A}_{BC}=\delta^{A}_{B}\lambda,_{C}+
\frac{1}{D-2}(\sigma,_{B}\delta^{A}_{C}-\sigma,_{D}G_{BC}G^{AD})
\label{S2}
\end{equation}
where $\lambda$ is an arbitrary function, which appears due to the 
existence of the Einstein-Kaufman $\lambda$-symmetry. If we choose the 
gauge $\lambda=\sigma/(D-2)$, then the antisymmetric part of 
$\Sigma^{A}_{BC}$ disappears and we get finally
\begin{equation}
\Sigma^{A}_{BC}(\sigma)=\frac{1}{D-2}(\delta^{A}_{B}\sigma,_{C}+
\delta^{A}_{C}\sigma,_{B}-\sigma,_{D}G_{BC}G^{AD})
\label{S3}
\end{equation}

In the presence of fermions, for the case $D=4$, in addition to the 
$\sigma$-dependent contribution to the connection(\ref{GAM2}), there is 
the usual fermionic 
contribution $\Sigma^{(f)A}_{BC}$ which does not depend on $\sigma$ 
(see for example Ref.\cite{FERM}). However, even in the presence of 
fermions we can use the $\lambda$-transformation since 
$L_{f}$ (see 
eq.(\ref{Lf})) is invariant under the $\lambda$-transformation. Due to 
this, the $\sigma$-dependent contribution to the antisymmetric part of 
$\Sigma^{A}_{BC}$ can be set to zero also here. Therefore 
we can write 
\begin{equation}
\Sigma^{A}_{BC}=\Sigma^{A}_{BC}(\sigma)+\Sigma^{(f)A}_{BC} 
\label{SF}
\end{equation}
where $\Sigma^{A}_{BC}(\sigma)$ is again defined by eq.(\ref{S3}).  

In the vacuum, the $\sigma$-contribution (\ref{S3}) to the connection 
can be eliminated 
by a conformal transformation of the metric\cite{BD} accompanied by a 
corresponding 
transformation of the fields defining the measure $\Phi$. Indeed, in the 
vacuum the action (\ref{Act1}),(\ref{L1}) is invariant under local 
transformations \begin{equation}
G_{AB}(x)=J^{-1}G^{\prime}_{AB}(x)
\label{ES1}
\end{equation}
\begin{equation}
\Phi(x)=J^{-1}(x)\Phi^{\prime}(x)
\label{ES2}
\end{equation}

For $J=\chi^{2/(D-2)}$ we get $\chi^{\prime}\equiv 1$, 
$\Sigma^{\prime A}_{BC}(\sigma)\equiv 0$ and $\Gamma^{\prime A}_{BC}=
\{ ^{A}_{BC}\}^{\prime}$, where 
$\{ ^{A}_{BC}\}^{\prime}$
 are the Christoffel's coefficients corresponding 
to the new metric $G^{\prime}_{AB}$. The appropriate generalization of 
the local symmetry (\ref{ES1}),(\ref{ES2}) in the presence of fermions 
will be discussed in Sec. 5. The extension of applicability of this 
local symmetry for realistic matter models will be discussed in Sec.6.

For the case where the measure $\Phi$ is given by eq.(\ref{Fi}), the 
transformation
(\ref{ES2})  
can be the result of a diffeomorphism
$\varphi_{a}\longrightarrow\varphi^{\prime}_{a}=
\varphi^{\prime}_{a}(\varphi_{b})$ in the space of the scalar fields 
$\varphi_{a}$ (see Ref.\cite{GK2}). Then $J=
Det(\frac{\partial\varphi^{\prime}_{a}}{\partial\varphi_{b}})$.

If we take the choice (\ref{MesA}), then eq.(\ref{ES2}) for a given $J$ 
may be interpreted as the result of the gauge transformation (\ref{BS}). 
\bigskip

\section{Equations of motion}

\bigskip

First we study equations that originate from the variation with respect 
to the measure fields. If the measure is defined using the antisymmetric 
tensor field $A_{A_{2}\ldots A_{D}}$ as the dynamical variable, we obtain
\begin{equation}
\epsilon^{A_{1}\ldots 
A_{D}}\partial_{A_{D}}\lbrack -\frac{1}{\kappa}R(\Gamma,G)+ L_{m}\rbrack =0
\label{AEM}
\end{equation}
which means that
\begin{equation}
-\frac{1}{\kappa}R(\Gamma,G)+L_{m}=M=constant
\label{1Int}
\end{equation}

If we consider the case where the measure is defined as in eq.(\ref{Fi}), 
we obtain instead of (\ref{AEM}), the equation
\begin{equation}
A^{B}_{b}\partial_{B}\lbrack -\frac{1}{\kappa}R(\Gamma,G)+
L_{m}\rbrack =0
\label{FEM}
\end{equation}
where $A^{B}_{b}=\varepsilon_{a_{1}\ldots a_{D-1}b}
\varepsilon^{A_{1}\ldots A_{D-1}B}
(\partial_{A_{1}}\varphi_{a_{1}}) \ldots
(\partial_{A_{D-1}}\varphi_{a_{D-1}})$. Since 
$A_{b}^{A}\partial_{A}\varphi_{b^{\prime}}=D^{-1}\delta_{bb^{\prime}}\Phi$
it follows that  $Det (A_{b}^{A}) = 
\frac{D^{-D}}{D!}\Phi^{D-1}$, so that if $\Phi\neq 0$, eq.(\ref{1Int}) is 
again obtained.

Therefore the two approaches for defining the measure which implements 
the NGVE principle, give, under regular conditions, the same equation 
(that is eq.(\ref{1Int}) ). The case where the measure is defined as in 
eq.(\ref{Fi}), provides with an extra possibility, which is that 
(\ref{1Int}) 
may not be satisfied if $\Phi=0$. That is one can envision a scenario 
where the integration constant $M$ in eq.(\ref{1Int}) could change while 
going through a singular surface with $\Phi=0$. This possibility and its 
cosmological consequences will be studied in a separate work.

Let us now study equations that originate from variation with respect to 
$G^{AB}$. For simplicity we present here the calculations for the case 
where there are no fermions. Performing the variation with respect to 
$G^{AB}$ we get
\begin{equation}
-\frac{1}{\kappa}R_{AB}(\Gamma)+\frac{\partial L}{\partial G^{AB}}=0
\label{RAB}
\end{equation}

Contracting eq.(\ref{RAB}) with $G^{AB}$ and making use eq.(\ref{1Int}) we 
get the constraint
\begin{equation}
G^{AB}\frac{\partial(L_{m}-M)}{\partial G^{AB}}-(L_{m}-M)=0
\label{con}
\end{equation}

This constraint has to be satisfied for all components (in the functional 
space) of the function $L_{m}$. In particular, for the constant part 
denoted $<L_{m}>$, which is relevant to a maximally symmetric vacuum state, 
we get
\begin{equation}
<L_{m}>-M=0
\label{SCCP}
\end{equation}

Inserting (\ref{SCCP}) in eq.(\ref{1Int}) we see that in the maximally 
symmetric vacuum the scalar curvature $R(\Gamma)$ is equal to zero. As 
we have seen in the previous section, the $\sigma$-contribution to the 
connection can be eliminated in the vacuum by  the
transformations (\ref{ES1}),(\ref{ES2}) (notice that due to the 
NGVE-principle, the constant part of the matter Lagrangian density 
$<L_{m}>$ does not alter the result that the action (\ref{Act1}) in the 
vacuum is invariant under the transformations (\ref{ES1}),(\ref{ES2}). This 
is because the measure $\Phi$ is a total derivative and therefore 
constant part of the Lagrangian density does not  contribute into 
equations of motion). Then in terms of the new metric $G^{\prime}_{AB}$, 
the scalar curvature $R(\Gamma,G)$ becomes the usual scalar curvature 
$R(G^{\prime}_{AB})$ of the Riemannian space-time with the metric 
$G^{\prime}_{AB}$. Therefore we conclude that the Riemannian scalar 
curvature vanishes in the maximally symmetric vacuum. In the presence of 
fermions the constraint (\ref{con}) has to be generalized. For more 
details about fermionic models see the next section.    

\bigskip

\section{Some matter models which satisfy automatically the constraint 
(\ref{con}) and local Einstein symmetry}

\bigskip

As we have seen, the consistency of the equations of motions demands the 
constraint (\ref{con}) to be satisfied. Here we are going to 
present theories where the constraint (\ref{con}) is associated with the 
existence of a local symmetry, which we have already identified in the 
vacuum case, i.e. the symmetry (\ref{ES1}),(\ref{ES2}), which is associated
with space-filling brane gauge invariance or with diffeomorphism 
invariance of the internal space of the fields $\varphi_{a}$. The model in 
the absence of this symmetry can also make sense\cite{GK2}, but then the 
geometrical interpretation of the theory is lost (in this case the 
constraint can still hold, but then the symmetry degrees of freedom 
becomes physical). Therefore, in what follows we will discuss cases when 
the local symmetry (\ref{ES1}),(\ref{ES2}) holds (possibly appropriately 
generalized) even when matter fields are introduced (we called this 
symmetry "local Einstein symmetry").

The  following examples satisfy the local Einstein symmetry  and 
constraint (\ref{con}) (the cases of gauge fields, massive scalar fields 
and massive fermions will be discussed in Sec.6). 

1.Scalar fields without potentials, including fields subjected to non
linear constraints, like the $\sigma$ model\cite{GK1},\cite{GK2}.  The 
general coordinate
invariant action for these cases has the form $S_{m} =\int L_{m}\Phi
d^{D}x$ where $L_{m}=
\frac{1}{2}\sigma,_{A}\sigma,_{B}g^{AB}$.

2.Matter consisting of fundamental bosonic strings\cite{GK1},\cite{GK2}.  
The constraint
(\ref{con}) can be verified by representing the string action in the
$D$-dimensional form where $G_{AB}$ plays the role of a background
metric.  For example, bosonic strings, according to our formulation,
where
the measure of integration in a $D$ dimensional space-time is chosen to
be $\Phi d^{D}x$, will be governed by an action of the form:
\begin{eqnarray}
S_{m}& =&\int
L_{string}\Phi d^Dx,\\
L_{string}&=& -T\int d\sigma
d\tau\frac{\delta^{D}(x-X(\sigma,\tau))}{\sqrt{-G}}
\sqrt{Det(G_{AB}X^{A}_{,a}X^{B}_{,b})}
        \label{String}
\end{eqnarray}
where $\int L_{string}\sqrt{-G}d^{D}x$ would be the action of a string
embedded
in a $D$-dimensional space-time in the standard theory; $a,b$ label
coordinates in the string world sheet and $T$ is the string tension.
Notice that under a transformation (\ref{ES1}),
$L_{string}\rightarrow J^{(D-2)/2}L_{string}$ , therefore
concluding that $L_{string}$ is a homogeneous function of $G^{AB}$
of degree one, that is constraint(\ref{con}) is satisfied only if $D=4$.

3.It is possible\cite{GK1},\cite{GK2} to formulate  the point particle 
model of matter in four dimensions ($D=4$) in a way
such that eq.(\ref{con}) is satisfied.  This is because for the
free falling point
particle a variety of actions are possible (and are equivalent in the
context of general relativity).  The usual actions in the 4-dimensional 
space-time with the metric $g_{\mu\nu}$ are taken to be $S=-m\int F(y)ds$, 
where $y=g_{\mu\nu}\frac{dX^{\mu}}{ds}\frac{dX^{\nu}}{ds}$ and $s$ is
determined to be an affine parameter except if $F=\sqrt{y}$, which is
the case of reparametrization invariance.  In our model we must take
$S_{m}=-m\int L_{part}\Phi d^{4}x$ with $L_{part}=
-m\int ds\frac{\delta^{4}(x-X(s))}{\sqrt{-g}}F(y(X(s)))$ where $\int 
L_{part}\sqrt{-g}d^{4}x$ would be the action of a point particle in 4
dimensions in the usual theory.  For the choice $F=y$, constraint
(\ref{con}) is satisfied.  Unlike the case of general relativity,
different choices of $F$ lead to unequivalent theories.

4.In the presence of Dirac fermions  with the Lagrangian density 
(\ref{Lf}) (in four
dimensions, $D=4$)
the local Einstein symmetry (\ref{ES1}),(\ref{ES2}) is appropriately 
generalized to
\begin{equation}
e^{a}_{\mu}(x)=J^{-1/2}(x)e^{\prime a}_{\mu}(x);\ e_{a}^{\mu}(x)=
J^{1/2}(x)e^{\prime\mu}_{a}(x)
\label{LA1}
\end{equation}
\begin{equation}
\Phi(x)=J^{-1}(x)\Phi^{\prime}(x)
\label{LA2}
\end{equation}
\begin{equation}
\psi(x)=J^{1/4}(x)\psi^{\prime}(x);\ 
\overline{\psi}(x)=J^{1/4}(x)\overline{\psi}^{\prime}(x) 
\label{LA3}
\end{equation}
provided that $V(\overline{\psi}\psi)\propto(\overline{\psi}\psi)$ or 
$(\overline{\psi}\gamma_{i}\psi)(\overline{\psi}\gamma^{i}\psi)$, which 
describe a Nambu - Jona-Lasinio type interaction\cite{NJL}.
Notice that in this case the condition for the invariance of the action with 
the matter Lagrangian 
(\ref{Lf}) under the transformations (\ref{LA1}-(\ref{LA3}) is not just 
the simple homogeneity of degree 1 in $g^{\mu\nu}$ or degree 2 in 
$e_{a}^{\mu}$, because of the presence of the fermion transformation 
(\ref{LA3}). However, the invariance under (\ref{LA1})-(\ref{LA3}) 
together with the fermionic equations of motion gives now the constraint 
in the form
\begin{equation}
e^{a\mu}\frac{\partial L_{m}}{\partial e^{a\mu}}-2L_{m}=0
\label{conf}
\end{equation}

This constraint was discussed in Ref.\cite{GK2} without reference to the 
generalized local Einstein symmetry (\ref{LA1})-(\ref{LA3}).
From the results of Sec.3 concerning  the $\lambda$-symmetry of the 
fermionic term of the action (see eq.(\ref{SF})) and making use the 
local Einstein symmetry (\ref{LA1})-(\ref{LA3}) we can   
reduce  the connection to  the 
usual one in the presence of fermions \cite{FERM} .

\bigskip

\section{Gauge fields, scalar fields with nontrivial potentials and 
massive fermions from a six dimensional  theory}

\bigskip

We have seen in the previous paper\cite{GK2} that when trying to 
introduce gauge fields into the theory in a way which is consistent with 
the local Einstein symmetry(\ref{ES1}),(\ref{ES2}), this runs against the 
problem that the gauge field kinetic energy $G^{AB}G^{CD}F_{AC}F_{BD}$ 
has  homogeneity of degree  $2$ in $G^{AB}$ instead of degree $1$ which is 
needed in order to satisfy the constraint (\ref{con}). We have shown also 
in Ref.\cite{GK2} how this problem can be avoided in the framework of the 
Kaluza-Klein approach. However, the solution of this problem suggested in 
Ref.\cite{GK2} seems to be not realistic enough.
  
We now will show how it is possible to construct more realistic models 
then those discussed before, by working in the context of a higher 
dimensional theory with two or more compactified dimensions with 
curvature. In this case we can introduce curvature dependence in 
prefactors of gauge field kinetic energy, scalar field potentials or 
fermionic mass such that the local Einstein symmetry be an exact symmetry.

In this case we consider an action of the form (the case of fermions will 
be considered at the end of this section)
 \begin{equation}
S=\int\Phi d^{6}x\lbrack-\frac{1}{\kappa}R(\Gamma,G)-
\frac{\lambda}{R(\Gamma,G)}F_{AB}F^{AB}+
\frac{1}{2}G^{AB}\partial_{A}\varphi\partial_{B}\varphi-
R(\Gamma,G)V(\varphi)\rbrack ,
\label{A6}
\end{equation}
where $F_{AB}\equiv \partial_{A}A_{B}-\partial_{B}A_{A}$.
The prefactors $\lambda/R(\Gamma,G)$ in the gauge field kinetic 
energy and $R(\Gamma,G)$ in the scalar field potential $V(\varphi)$ are 
required so 
as to preserve the local Einstein symmetry (\ref{ES1}),(\ref{ES2}).

The simplest realization of this idea is achieved in a six dimensional 
model where two dimensions are compactified into a sphere. We will see 
that the models we discuss allow and seem to prefer this type of 
compactification. Furthermore, for solutions which are maximally 
symmetric in the remaining four dimensions, the noncompactified 
4-dimensional space-time is only  Minkowski space. This means that 
starting from a higher dimensional model we achieve a four dimensional 
solution of the cosmological constant problem.

The simplest model that  respects the local Einstein symmetry and 
gives rise to $M^{4}\times S^{2}$ compactified solution, is a model where 
compactification is triggered by a non linear sigma model. 

In this case
\begin{equation}
S=\int\Phi d^{6}x\lbrack-\frac{1}{\kappa}R(\Gamma,G)+
\frac{1}{2}G^{AB}\partial_{A}\vec{\phi}\cdot\partial_{B}\vec{\phi}\rbrack
\label{GZC}
\end{equation} 
where the scalar field $\vec{\phi}$ is an isovector constrained to satisfy
$\vec{\phi}^{2}=f^{2}=constant$. This model is invariant under the local 
Einstein symmetry(\ref{ES1}),(\ref{ES2}).

For the hedgehog configuration
\begin{equation}
\vec{\phi}=f(\cos\theta,\ \sin\theta\sin\varphi,\ \sin\theta\cos\varphi) 
\label{hed}
\end{equation}
the $M^{4}\times S^{2}$ metric
\begin{equation}
ds^{2}=-dt^{2}+d\vec{x}^{2}+b^{2}(d\theta^{2}+\sin^{2}\theta d\varphi^{2})
\label{Mhed}
\end{equation}
where $b$ is an arbitrary constant, is a solution in the gauge $\chi =1$ 
(that is in the gauge where the gravitational equations coincide with the 
6-dimensional Einstein's equations) provided $f^{2}=2/\kappa$ 
(see Ref.\cite{Group}). If one wants to avoid the fine tuning of this 
parameter of the Lagrangian one can use instead a no scale non linear 
sigma model where the size of the surface in isospin space is determined 
dynamically\cite{G}). In this case $b$ is not determined by the equations 
of motion. The $M^{4}\times S^{2}$ form of compactification can be seen 
quite directly from the form of the equations $R_{AB}=
\kappa\frac{\partial L}{\partial G^{AB}}$, since for the case 
(\ref{hed}),(\ref{Mhed}) we immediately obtain the condition 
$R_{\mu\nu}=0$
($\mu,\nu=0,1,2,3$) and $R^{\varphi}_{\varphi}+R^{\theta}_{\theta}=R=
2/b^{2}\neq 0$. Finally, we should point out that the possible 
Kaluza-Klein gauge fields acquire a big mass in this case\cite{GZ}, so 
they don't appear in the low energy physics.

It is interesting to see that it is possible to induce $M^{4}\times S^{2}$ 
compactification from a gauge-field monopole configuration in the extra 
dimensions $S^{2}$. Let us consider the first two terms of the action 
(\ref{A6}), i.e. the  action describing gravity 
+ gauge fields in a locally Einstein symmetric way. For the magnetic 
monopole $A_{A}=0$ if $A=0,1,2,3$,\ $A_{\theta}=0$,\ $A_{\varphi}=
m(\cos\theta\mp 1)$ and if
\begin{eqnarray}
ds^{2}=g_{\mu\nu}dx^{\mu}dx^{\nu}+ b^{2}(x)d\Omega^{2}
\nonumber\\
d\Omega^{2}=d\theta^{2}+\sin^{2}\theta d\varphi^{2},
\nonumber\\
\mu,\nu =0,1,2,3,
\label{MS6}
\end{eqnarray}
we find the equations
\begin{equation}
\frac{1}{\kappa}R_{\mu\nu}(\Gamma)=
\lambda\frac{R_{\mu\nu}(\Gamma)}{R^{2}(\Gamma,G)}F^{2}
\label{E64}
\end{equation}
\begin{equation}
\frac{1}{\kappa}R_{ab}(\Gamma)=
\lambda\frac{R_{ab}(\Gamma)}{R^{2}(\Gamma,G)}F^{2}-
\frac{2\lambda}{R(\Gamma,G)}F_{ac}F^{c}_{b},\ (a,b=\theta,\varphi)
\label{E62}
\end{equation}
where $F^{2}=F_{AB}F^{AB}$.

If $R_{\mu\nu}(\Gamma)\neq 0$, then eqs.(\ref{E64}),(\ref{E62}) imply 
$F^{2}=0$, which is not consistent with the monopole ansatz. Using that 
\begin{equation}
R_{\mu\nu}(\Gamma)=0,
\label{64f}
\end{equation}
we see that $R(\Gamma,G)=R^{\theta}_{\theta}+R^{\varphi}_{\varphi}$ and 
from eq.(\ref{E62}) we get 
\begin{equation}
R^{2}(\Gamma,G)=-\kappa\lambda F^{2}.
\label{F2}
\end{equation}

Notice that the action (\ref{A6}) respects the $\lambda$-symmetry. Due to 
the 
local Einstein symmetry of the action, we can again fix the gauge where 
$\chi\equiv 1$ (that is $\Phi\equiv\sqrt{-g}$). Then the 
$\sigma$-contribution (\ref{S3}) to the connection is equal to zero.

When working with the action of Sec.3, we were able to find the 
connection $\Gamma^{A}_{BC}$ as a solution of eq.(\ref{GAM1}) without 
using the equations of motion which follow from the variation  with 
respect to $G^{AB}$. Now however the scalar curvature enters in the 
equation obtained from the variation with respect to $\Gamma^{A}_{BC}$. 
Therefore we have to solve these equations together. 

We are interested now in solutions which are maximally symmetric with 
respect to the remaining four dimensions. Therefore in (\ref{MS6}) we 
choose $g_{\mu\nu}(x)$ as a metric of a maximally symmetric 4-dimensional 
space-time with 10 Killing vectors and $b(x)=constant$. Then $F^{2}$ is a 
constant. In this case, the 
variation with respect to $\Gamma^{A}_{BC}$ leads again to an equation 
like (\ref{GAM1}) with a common factor $(-\frac{1}{\kappa}+
\frac{\lambda}{R^{2}(\Gamma,G)}F^{2})$. It follows from  
eq. (\ref{F2}), that this 
factor is not equal to zero. Therefore, in the case under consideration, 
the magnetic monopole configuration does not contribute to the 
connection and hence the solution of eq.(\ref{GAM1}) is now just the 
Christoffel's connection coefficients: $\Gamma^{A}_{BC}=
\{ ^{A}_{BC}\}$. It means that $R_{AB}(\Gamma)$ is just 
the usual Ricci tensor $R_{AB}(G_{CD})$ . Therefore eq.(\ref{64f}) 
takes the form 
$R_{\mu\nu}(g_{\lambda\sigma})=0$ where 
$R_{\mu\nu}(g_{\lambda\sigma})$ is the usual Ricci 
tensor of the four dimensional space-time with a maximally symmetric metric 
$g_{\lambda\sigma}$.
We conclude that {\em the very curved extra dimensions are necessarily 
accompanied only with the flat maximally symmetric four dimensional 
space-time, that is with Minkowski space}. 

 Equation (\ref{F2}) with 
$R=R(G_{AB})$ 
gives us the value of the strength of the magnetic monopole: 
$m=\sqrt{2/\kappa\lambda}$. Notice that the constant size $b$ of the 
extra dimensions is not determined. The existence of this flat direction 
is associated (from the 4-dimensional point of view) with a massless 
scalar field. Whether this is a phenomenological problem depends on the 
coupling and possible cosmological evolution of such scalar field. This 
will be studied in the future.

In this model there is no mass generation for the Kaluza-Klein gauge 
fields, which can therefore play a role in the low energy physics. It is 
also interesting to notice that what matter does in the extra dimensions 
produces directly curvature only in the extra dimensional space, at least 
in the ground state. That is, 
there is no mixing between Planck scale physics and low energy physics. 

This can be compared with the well known Freund-Rubin compactification, 
for example when applied to 
11-dimensional supergravity\cite{FR}, where an expectation value of a four 
index field strength $F_{ABCD}$ in four dimensions is responsible for curving 
four dimensions into an anti de Sitter space and also for the 
compactification of seven dimensions into a sphere, i.e., a complete mix 
up of the physics of compactification and the physics that dominates the 
large scale structure of the observed four dimensions.

With the addition of a potential of the form $R(\Gamma,G)V(\varphi)$ 
(consistent with 
the local Einstein symmetry) to eq.(\ref{A6}), the possibility of a scalar 
field with non 
trivial dynamics and in particular the possibility of mass for the scalar 
field and of spontaneous symmetry breaking appears in a straightforward 
fashion. Notice that phase transitions associated with a change of 
$<V(\phi)>$ correspond to a change in the effective Newton constant and 
not related to a change of vacuum energies, which cannot enter into the 
theory anyway.

Let us consider now the possibility of fermionic mass generation in 
4-dimensional space-time in a way 
consistent with the local Einstein symmetry. If we look for a term which 
generates a fermionic mass term as a result of compactification in the 
form $fR^{n}\overline{\psi}\psi  $ with dimensionless coupling constant 
$f$ , then $n$ has to be equal to $ 1/2$. In this case the only space-time 
dimension $D$  which allows the local Einstein symmetry 
(\ref{LA1})-(\ref{LA3}) is $D=6$. Therefore for the fermionic mass 
generation in 4-dimensional space-time (without introduction new 
dimensionful coupling constant) we have to start from the 6-dimensional 
model with the action
\begin{equation}
S_{f}=-\frac{i}{2}\int\Phi 
d^{6}x\lbrack(\nabla_{A}\overline{\psi})\gamma^{A}\psi-
\overline{\psi}\gamma^{A}\nabla_{A}\psi 
+2i\sqrt{R(\Gamma,G)}\overline{\psi}\psi\rbrack 
\label{Lmf}
\end{equation}
For $\gamma$-matrices and other quantities in six dimensions see 
Ref.\cite{Sal}. After the compactification of two extra dimensions into a 
sphere, the curvature of the sphere induces a mass for fermions in 
4-dimensional space-time.

\bigskip

\section{Discussion}
\bigskip

We have shown that the NGVE-principle in the context of the first order 
formalism solves the cosmological constant problem. In this paper we have 
formulated several models (in the above framework) that respect the local 
Einstein symmetry which has nice geometrical interpretations. Furthermore, 
in models where the local Einstein symmetry is the exact symmetry, we 
always have both constraint(\ref{con}) and possibility to obtain the 
measure $\sqrt{-G}$ by setting the gauge $\Phi=\sqrt{-G}$.

Using higher dimensional ($D\geq 6$) models  it is 
possible to maintain 
this local Einstein symmetry while constructing realistic models which 
allow for gauge fields, mass generation, spontaneous symmetry breaking,
 etc..  This is possible to realize in the presence of compactification 
of extra dimensions into a sphere and simultaneously achieving zero 
4-dimensional cosmological constant. This result is related to the fact 
that in such a model the physics that is responsible for compactification 
does not affect the geometry of the large scale structure of the 
uncompactified four dimensional space-time. 

Furthermore, in the case where we use the gauge model (\ref{A6}), 
compactification appears not only as a choice, since the alternative 
six-dimensional maximally symmetric vacuum with $R=0$ would be a sick 
vacuum. This is not only because $1/R$ is undefined but also because the 
small perturbations bring us to the region where the gauge field kinetic 
term has wrong sign which is of course an unstable regime.

Finally, if it is the case that the local Einstein symmetry can be 
maintained even after quantum corrections are considered, we get an 
interesting constraint on the form of the possible quantum corrections 
which can be only terms homogeneous of degree 1 in $G^{AB}$ like 
 for example $R_{AB}R^{AB}/R$.

 \section{ Acknowledgements}

We would like to thank N.Kaloper for interesting conversations.

\pagebreak

\end{document}